\documentclass[prl,amsmath,twocolumn,showpacs,superscriptaddress]{revtex4-1}
\usepackage{epsfig}

\newcommand{\noi}{\noindent}
\newcommand{\be}{\begin{equation}}
\newcommand{\ee}{\end{equation}}

\begin{document}

\title{Stationary growth and unique invariant harmonic measure of cylindrical DLA}
      
\author{Riccardo Marchetti}
\affiliation{La Sapienza Universit\`a di Roma,
Dipartimento di Fisica, P.le A. Moro 5, 00185 Rome, Italy.}                                         
\author{Alessandro Taloni}
\affiliation{La Sapienza Universit\`a di Roma,
Dipartimento di Fisica, P.le A. Moro 5, 00185 Rome, Italy.}                                         
\affiliation{CNR-IENI, Via R. Cozzi 53, 20125 Milano, Italy}
\author{Emanuele Caglioti}
\affiliation{La Sapienza Universit\`a di Roma, Dipartimento di Matematica
P.le A. Moro 5, 00185 Rome, Italy}
\author{Vittorio Loreto}
\affiliation{La Sapienza Universit\`a di Roma,
Dipartimento di Fisica, P.le A. Moro 5, 00185 Rome, Italy.}                                         
\affiliation{ISI foundation, Via Alassio 11/c
10126 Torino, Italy}
\author{Luciano Pietronero}
\affiliation{La Sapienza Universit\`a di Roma,
Dipartimento di Fisica, P.le A. Moro 5, 00185 Rome, Italy.}                                     
\affiliation{CNR-ISC, Via dei Taurini 19, 00185 Rome, Italy}
\pacs{}

\begin{abstract}
We prove that the harmonic measure is stationary, unique and invariant on the interface of DLA growing on a cylinder surface. We provide a detailed theoretical analysis puzzling together multiscaling, multifractality and conformal invariance, supported by extensive numerical simulations of clusters built using conformal mappings and on lattice. The growth properties of the active and frozen zones are clearly elucidated. We show that the  unique scaling exponent characterizing the stationary growth is the DLA fractal dimension.  
\end{abstract}
\maketitle

%
%
{\em Introduction.--} 
Albeit the diffusion limited aggregation model (DLA) was introduced 30 years ago ~\cite{Witten-Sander}, it still embodies the perfect puzzle for theorists. In spite of the enormous amount of numerical works, only few rigorous results are proven, and these all concern the \emph{generalized dimensions}  of the aggregate. The notion of varying dimensionality characterizes the DLA growth  in two ways: multiscaling and multifractality. Multiscaling suggests that the aggregate's fractal dimension attains different local values $d(r/R)$ on each circular ring of radius $r$, while the overall gyration radius  $R(N)\sim N^{1/D}$  ($N$  and $D$ cluster's particles number  and  fractal dimension) ~\cite{Coniglio,Amitrano}. 
Multifractality ~\cite{multifractality}  entails the introduction of  generalized dimensions $D(q)$, which correspond to the fractal dimension of the $q$ points correlation function. The aforementioned rigorous results are: $D(0)\geq 3/2$ ~\cite{Kesten}, $D(1)=1$ ~\cite{Makarov} and $D(3)=D(0)/2$ ~\cite{Halsey}. A simpler definition of $D(q)$ is provided by the connection with the harmonic measure, i.e.  the growth probability at the interface. In general, the determination of the harmonic measure within the fjords of the fractal, is a quite impossible task if resorting to the usual numerical techniques ~\cite{harmonic_numerical}. Iterated conformal mapping ~\cite{Hastings} provides a solution to this problem ~\cite{Jensen}: indeed it  is based on the representation of the growth dynamics via the convolution of elementary complex functions $\phi_{\lambda_n,\theta_n}(w)$, hereafter named  $\phi_{n}(w)$, which map the unitary circle onto a unitary circle with a bump of linear size $\sqrt{\lambda_n}$ placed at position $w=e^{i\theta_n}$ ~\cite{Davidovitch}. The ensuing function $z=\Phi^{(N)}\left(e^{i\vartheta}\right)=\phi_{1}
\circ\phi_{2}
\circ\cdots\circ\phi_{N}\left(e^{i\vartheta}\right)$ transforms the unitary circle $e^{i\vartheta}$ onto the cluster's interface $z$. Furthermore conformal mapping has been extended to DLA growing on a cylinder surface ~\cite{Taloni}: in this case the mapping function is 

\be
\Xi^{(N)}\left(e^{i\vartheta}\right)=\overline{-i\frac{L}{2\pi}\ln\left[\Phi^{(N)}\left(e^{i\vartheta}\right)\right]},
\label{mapping_cil}
\ee

\noi where $L$ is the cylinder circumference and $\Phi^{(N)}$ represents the radial deformation of the cylindrical aggregate (see  section I in suppl. mat.). The DLA overall height grows as $h(N)\simeq\frac{L}{2\pi}\ln F^{(N)}_1$ where $F^{(N)}_1=\Pi_{n=1}^N\sqrt{1+\lambda_n}$ is the first Laurent coefficient of $\Phi^{(N)}$ ~\cite{Hastings, Davidovitch}.

In this letter we provide another rigorous result concerning the harmonic measure of a DLA growing in a  cylindrical geometry. We show that the growth probability at the interface is stationary, unique and invariant, leading to the conformal invariance of the complex mapping function. Our framework provides the natural connection between multiscaling and multifractality, and offers a clearcut definition of the frozen and active zone of the aggregate. Moreover, we show that the stationarity leads to the appeareance of  the fractal dimension $D$ as the unique exponent governing the growth dynamics, in stricking contrast to  DLA  in radial geometry.

\begin{figure}
\begin{center}
\epsfig{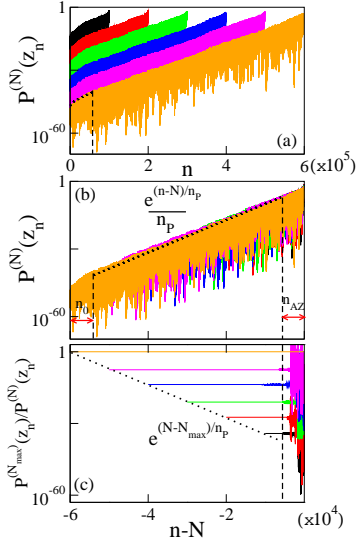}
\end{center}
\caption{
(Color online) Probability distribution on DLA interface.  $P^{(N)}(z_n)$ is shown for a single DLA realization ($\lambda_0=10^{-5}$, $L=1$) for $N=10^4,2\times 10^4, 3\times10^4, 4\times 10^4, 5\times 10^4$ and $6\times 10^4$ (from left to right). (a) $P^{(N)}(z_n)$ as function of generation $n$: dotted line corresponds to the self-affine regime, i.e. $\frac{n^\beta}{n_P^{\beta+1}} e^{-\frac{N-n}{n_P}}$ ($\beta=4/3$). (b) Rescaling of $P^{(N)}(z_n)$:  probabilities at different $N$ collapse on top of each other exhibiting the $N-n$ dependence. Dashed lines show the self-affine region ($n_0$) and the active zone ($n_{AZ}$). (c) Ratio between  $P^{(N_{max})}(z_n)$ ($N_{max}=6\times 10^4$) and the  curves in panel (a), showing the stationarity in the frozen zone.}
\label{fig1}
\end{figure}

{\em Height's scaling and growth velocity.--} The average height of a cylindrical aggregate grows according to the following law ~\cite{Taloni}: $\langle h(N)\rangle=L \left(\frac{N}{\langle n_0\rangle}\right)^\beta$ if $N\ll \langle n_0\rangle$ and $\langle h(N)\rangle=L \frac{N}{\langle n_0\rangle}$ if $N\gg \langle n_0\rangle$ ($\beta\simeq 4/3$, $\langle n_0\rangle=\frac{1}{\pi}\left(\frac{\sqrt{\lambda_0}}{L}\right)^{-D}$). The first regime accounts for a self-affine initial growth, while the DLA attains the linear self-similar regime after an average \emph{ time} $\langle n_0\rangle$, namely the transient on which the cluster forgets about its initial condition, i.e. the cylinder's baseline.
From the scaling of $\langle h(N)\rangle$ and the expression of $F^{(N)}_1$, one obtains $\langle h(N)\rangle\simeq\frac{L}{4\pi}\sum_{n=1}^{N}\langle\lambda_n\rangle$, which furnishes the expression of the average  height's growth velocity: $d\langle h(N)\rangle/dN\simeq\frac{L}{4\pi}\langle\lambda_N\rangle$. Thus the average elementary increment $\langle\lambda_n\rangle$ grows as $\sim 4\pi\beta \frac{n^{\beta-1}}{\langle n_0\rangle^\beta}$ for $n\ll \langle n_0\rangle$, and it gets to the stationary value $\frac{4\pi}{\langle n_0\rangle}$ for $n\gg \langle n_0\rangle$. This suggests that self-similarity is intimately connected to the height's velocity stationary growth, in contrast with DLA in radial geometry ~\cite{Davidovitch}: in this case the radius growth is clearly non-stationary and $\langle\lambda_n\rangle=2/(nD_R)$, with $D_R\simeq 1.71$  radial cluster's dimension.

\begin{figure}
\begin{center}
\epsfig{figure=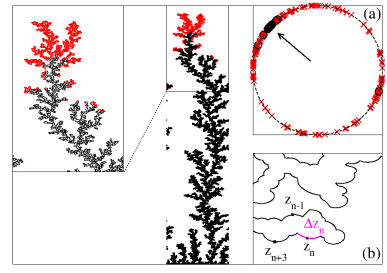, width=0.28\textwidth}
\end{center}
\caption{(Color online) Active zone of the DLA. Typical DLA realization obtained through (\ref{mapping_cil}) ($\lambda_0=10^{-4}, \,L=1$): black and red regions represent the frozen and the active zone. Inset (a): DLA interface on the unitary circle. Red crosses are the  boundaries of $\Delta\vartheta_n^N$ for $n\in[N-n_{AZ},N]$, black circles for $ n\ll N-n_{AZ}$: arrow shows where the counterimage of the frozen zone is mostly concentrated. Inset (b): boundary of the DLA showing $\Delta z_n$ (magenta)  and the representative point $z_n$.}
\label{fig2}
\end{figure}

{\em Harmonic measure--} We now proceed to the evaluation of the harmonic measure ~\cite{Witten-Sander}.   The DLA interface is  the union of different arcs, which are the (remaining) boundaries of the $N$ particles  composing the cluster. Each arc may be labeled by $\Delta z_n$ where $n$ is the generation when the $n$-th bump was added to the structure; besides, we identify the middle point $z_n$ as the representative point of the entire arc (Fig.[\ref{fig2}], inset (b)), and we calculate the probability $P^{(N)}(z_n)$. The latter is the normalized electric field on the fractal's interface ~\cite{Niemeyer}, i.e. $P^{(N)}(z_n)\propto \left|E^{(N)}(z_n)\right|$, which  is connected to the Jacobian of (\ref{mapping_cil}) as $ \left|E^{(N)}(z_n)\right|=\left|\nabla\Xi^{(N)}\left(e^{i\vartheta_n^N}\right)\right|^{-1}$ ~\cite{Hastings,Davidovitch}. The angle $\vartheta_n^N$ represents the counterimage of  $z_n$ on the unitary circle, and must be subjected to a reparametrization whenever a new particle is added ~\cite{Jensen,Taloni}. Indeed consider a DLA at two different generations $N$ and $N-k$, with $k\in[1,N-n]$: the position $z_n$ remains unchanged whether the cluster has $N$ or $N-k$ particles, i.e. $z_n=\Xi^{(N)}\left(e^{i\vartheta_n^N}\right)=\Xi^{(N-k)}\left(e^{i\vartheta_n^{N-k}}\right)$, yielding

\be
e^{i\vartheta_n^{N}}=\phi^{-1}_{N}\circ\phi^{-1}_{N-1}\circ\cdots\circ\phi^{-1}_{N-k+1}\left(e^{i\vartheta_n^{N-k}}\right),
\label{branch_cuts}
\ee

\noi  where $\phi^{-1}_{n}$ is the inverse of  $\phi_{n}$ ~\cite{Jensen,Taloni}. Thus any $\vartheta_n^{N}$   can be determined from its initial value $\vartheta_n^{n}\simeq\theta_n$. The electric field at $z_n$ can be calculated as $\left|E^{(N)}(z_n)\right|=\frac{\left|E^{(n)}(z_n)\right|}{\Pi_{k=n+1}^N\left|\phi'_{k}\left(e^{i\vartheta_n^{k}}\right)\right|}$
\noi where $\phi'_{n}$ is the derivative of $\phi_{n}$. Fig.[\ref{fig1}] shows the probability $P^{(N)}(z_n)$ as a function of $n$, for a single DLA realization at 6 different values of $N$. The harmonic measure $\mu_{N,n}$ is the ensemble average of  $P^{(N)}(z_n)$, i.e. $\mu_{N,n}\left(\frac{\sqrt{\lambda_0}}{L}\right)=\langle P^{(N)}(z_n)\rangle$;  from the numerics its behavior is stationary, depending solely on the difference $N-n$:

\be
\mu_{N,n}\simeq\frac{1}{\langle n_P\rangle}\times\left\{
\begin{array}{ccc}
\left(\frac{n}{\langle n_P\rangle}\right)^\beta e^{-\frac{N-n}{\langle n_P\rangle}} &  & n\ll \langle n_0\rangle\\
 e^{-\frac{N-n}{\langle n_P\rangle}}&  & \langle n_0\rangle \ll n\ll N-\langle n_{AZ}\rangle\\
\left(\frac{\sqrt{\lambda_0}}{L}\right)^{\alpha\left(\frac{n-N}{\langle n_P\rangle}\right)} &  & N-\langle n_{AZ}\rangle\ll n\ll N
\end{array}\right.
\label{harmonic_meas}.
\ee

\begin{figure}
\begin{center}
\epsfig{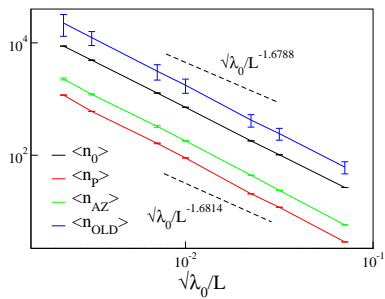}
\end{center}
\caption{(Color online) Scaling of the characteristic times ruling the DLA growth. Different characteristic times seems to fulfill  the scaling law $\sim\left(\frac{\sqrt\lambda_0}{L}\right)^{-D}$,  with $D\simeq 1.67$ fractal dimension. We took $\langle n_0\rangle=\frac{4\pi}{\langle \lambda_n\rangle}$ and $L=1$.}
\label{fig3}
\end{figure}

\noi   $n_{AZ}$ represents the number of particles that compose the \emph{active zone} of the DLA ~\cite{Plischke}. Firstly,  we focus on the \emph{frozen zone}  for which $n\ll N-\langle n_{AZ}\rangle$: a quantitative analysis  of the  measure's scaling in  the active region, will be given in the following.  The probability of a point belonging to the frozen interface exhibits an apparent exponential decay, with a characteristic \emph{time} $n_P$. This arises from the  electric field expression, indeed for any point in the frozen zone $\left|\phi'_{k}\left(e^{i\vartheta_n^{k}}\right)\right|\simeq e^{1/n_P}$, if  $k\gg n+n_{AZ}$. This suggests that a point in the frozen zone cannot influence the growth dynamics, since the probability in $z_n$ does not depend on the specific choice of the elementary function $\phi_{k}$ when $k\gg n+n_{AZ}$. On the other side, it points to the notion of \emph{conformal invariance} ~\cite{Hastings}, since two conformal transformations $\phi_{n}$ and $\phi_{k}$ commute whenever $k\gg n+n_{AZ}$. Indeed, given a mapping function $\Xi^{N}(w)$ with $N-n_{AZ}\gg k\gg n+n_{AZ}$, the size of the new bump $\sqrt{\lambda_{N+1}}$ and the ensuing growth dynamics will remain unchanged whether $\phi_{k}$ is swapped with $\phi_{n}$.  $n_P$ can be accurately measured for any $z_n$ lying in the frozen zone, indeed from (\ref{harmonic_meas}) one has $ n_P=(N_2-N_1)/\left[\ln P^{(N_1)}(z_n)-\ln P^{(N_2)}(z_n)\right]$, with $N_2$ and $N_1$  two arbitrary generations. The scaling of  $\langle n_P\rangle$  is displayed in Fig.[\ref{fig3}].

{\em Active zone--} A zoom of $\mu_{N,n}$ for $n\in[N-\langle n_{AZ}\rangle,N]$  is shown in suppl. mat. (Fig.[9] panel (a)). The active zone is  the region where new particles join the existing cluster ~\cite{Plischke, Meakin, Ossadnik}; our  observation indicates that it corresponds to the region occupied by the last $n_{AZ}$ (Fig.[\ref{fig2}]). Conformal mapping transforms $\mu_{N,n}$ to the uniform measure on the unitary circle ~\cite{Hastings,Davidovitch}:

\be
P^{(N)}(z_n)dz_n=\frac{d\vartheta_n^N}{2\pi}
\label{measure_map}
\ee

\noi where $dz_n$ represents the infinitesimal interface's portion  around  $z_n$. Hence, we can  define the  active zone  through the relation $\sum_{n=N-n_{AZ}}^{N}\frac{\Delta\vartheta_n^N}{2\pi}\geq 0.95$ ( Fig.[\ref{fig2}] inset (a)). The  scaling of $\langle n_{AZ}\rangle$ is shown in Fig.[\ref{fig3}].
Now, a well-established fact is that $\mu_{N,n}$  in the active zone exhibites a multifractal scaling ~\cite{multifractality,Jensen,Jensen_2}. In this context, growth probability scales differently in different regions characterized by the multifractal exponent $\alpha$, namely $\mu\left(\frac{\sqrt{\lambda_0}}{L}\right)\sim \left(\frac{\sqrt{\lambda_0}}{L}\right)^{\alpha}$ ($\alpha_{min}\leq \alpha\leq \alpha_{max}$) ~\cite{Jensen_2}. Numerical simulations show that $\alpha$ exhibites a stationary dependence on the ratio $(n-N)/\langle n_P\rangle$: $\alpha\left(\frac{n-N}{\langle n_P\rangle}\right)$ (Fig.[9] panel (b) in suppl. mat.). This relation bridges together multiscaling and multifractality. Indeed, it has been proposed ~\cite{Coniglio} that, although distinct phenomena, multiscaling and multifractality may provide an equivalent description whether $\alpha=\alpha\left(\frac{r}{R}\right)$, where $r$ and $R$ represent an inner and the overall radius of a radial cluster. Since the average radii of the radial deformation of the aggregate scale as $\langle r\rangle\sim e^{\frac{2\pi n}{\langle n_0\rangle}}$ and $\langle R\rangle\sim e^{\frac{2\pi N}{\langle n_0\rangle}}$  (Fig.[1] in suppl. mat.),  we get  $\alpha\left(\frac{r}{R}\right)\equiv\alpha\left(\frac{n-N}{\langle n_P\rangle}\right)$ where $\langle n_P\rangle\sim \langle n_0\rangle$ (Fig.[\ref{fig3}]). However this relationship does not hold for truly radial DLA.

\noi At this point, a question arises:  how long does  $z_n$  take to pass from the active  to the frozen zone?   This question is better addressed on the unitary circle. Indeed, when a new bump is created at $\theta_n\simeq \vartheta_n^n$, its size on the unitary circle is approximately $\sqrt{\lambda_n}\simeq\Delta\vartheta_n^n$  ~\cite{Davidovitch}, with  $\langle \Delta\vartheta_n^n\rangle\simeq\sqrt{\frac{4\pi}{\langle n_0\rangle}}$. When $N>n$, the angle $\vartheta_n^N$ changes its position due to (\ref{branch_cuts}), and  $\Delta\vartheta_n^N$ shrinks because of the decay of the probability $P^{(N)}(z_n)$ (\ref{measure_map}). Eventually $\Delta\vartheta_n^N$ shrinks to zero (within the machine precision) and  the angles $\vartheta_n^N$, counterimages of the points lying in the frozen zone, become indistinguishable on the unitary circle (Fig.[\ref{fig2}] inset (a)). We measured the average time $\langle n_{OLD}\rangle$ for which $\Delta\vartheta_n^{n+n_{OLD}}\simeq 0$ (Fig.[\ref{fig3}]). Moreover the fact that the frozen zone almost corresponds to a unique angle on the unitary circle, explains why 
$\left|\phi'_{k}\left(e^{i\vartheta_n^{k}}\right)\right|\simeq\left|\phi'_{k}\left(e^{i\vartheta_{n'}^{k}}\right)\right|\simeq e^{1/n_P}$ ($k\gg n^{(')}+n_{AZ}\sim n^{(')}+n_{OLD}$) and the ensuing conformal invariance, as any $\phi_{k}\left(e^{i\theta_{N}}\right)$ commutes with $\phi_{n}\left(e^{i\theta_{N}}\right)$ leaving unaffected the value of $\Xi^{(N)}\left(e^{i\theta_{N}}\right)$ ($N-n_{OLD}\gg k\gg n+n_{OLD}$).

{\em DLA collapse--} So far, we provided a strong evidence of the stationarity of the harmonic measure. The next step is to prove that it is unique and invariant: indeed, estimates of $\mu_{N,n}$ could be strongly affected by fluctuations and/or memory effects.  In general, a sufficient condition for the assessment if a stochastic process has one
invariant measure, is the coupling of two realizations of the process
with different initial conditions but same randomness.
If one is able to prove that both
processes collapse with probability one,
this means that there exists a
unique invariant measure \cite{MaSc} (see section II in suppl. mat.).

\noi In conformal mapping, the stochastic process is defined by the angles $\left[\theta_1,\ldots,\theta_N\right]$. Hence, we consider as independent initial conditions $\left[\theta_1^{(1)},\ldots,\theta_{N_{init}}^{(1)}\right]$ and $\left[\theta_1^{(2)},\ldots,\theta_{N_{init}}^{(2)}\right]$  ($N_{init}\gg \langle n_0\rangle$), and we extract subsequent  angles according to  $\theta_{N}^{(1)}=\theta_{N}^{(2)}$ ($N> N_{init}$):  collapse arises if and only if  $_{(1)}\Delta\vartheta_n^{N}=$ $_{(2)}\Delta\vartheta_n^{N}$, $\forall n\in[N-\langle n_{AZ}\rangle,N]$ (Fig.[5] in suppl. mat.).  However, this procedure is strongly affected by sistemical errors induced by  $\phi_n$  ~\cite{Davidovitch,Taloni,Stepanov} (section III of suppl. mat): if a growth attempt is made close to the frozen regions, unphysical particles tend to fill the fjords of the aggregates, leading to clusters' divergence rather than collapse. Thus, we apply the collapsing procedure to DLA on lattice.

\begin{figure}
\begin{center}
\epsfig{figure=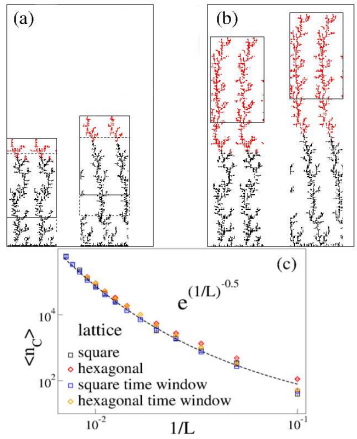, width=0.26\textwidth}
\end{center}
\caption{(Color online) DLA collapse. Collapsing protocols for DLA on lattice (lattice unit = 1). (a) Different initial conditions: two DLA grown independently (black particles); red particles follow the same Brownian paths on the cylinder surfaces. Frames represent the spatial window ($3L$) within which collapse can arise. (b) Collapsed DLA. (c) Collapsing time $\langle n_C\rangle $ obtained with two different collapsing protocols in both square (square symbols) and hexagonal lattice (diamonds):  spatial window and time window collapse criteria give $e^{(1/L)^{-0.5}}$.
}
\label{fig4}
\end{figure}

\noi For DLA on lattice, randomness is given by the Brownian nature of the upcoming particle's path: a collapsing protocol may consist on taking the same diffusive trajectories, for particles released from the upper  cylinders' boundaries in both DLA. Therefore, after two initial conditions have been built  (Fig.[\ref{fig4}] panel (a)), protocol is started and it is stopped only when both DLA have collapsed (Fig.[\ref{fig4}] panel (b)), i.e. when they are identical within a window of height $3L$. The average collapsing time $\langle n_C\rangle$ shows a stretched exponential  behaviour, $\langle n_C\rangle\sim e^{(1/L)^{-0.5}}$,  valid in square and in hexagonal lattices (Fig.[\ref{fig4}] panel (c)). The outlined collapsing protocol requires that both DLA are overlapping in a \emph{spatial} window: we now want that collapse occurs when they are identical within a \emph{temporal} window $\langle n_{AZ}\rangle$. In this case, DLA collapse is the physical distance between the  positions of the last $\langle n_{AZ}\rangle$ homologous particles is 0. Our results strongly indicate that the stretched exponential decay  is robust to the change of lattice geometry and definition of collapsing criteria (Fig.[\ref{fig4}] panel (c)), but is definitely different from what has been observed for $\langle n_0\rangle$, $\langle n_P\rangle$, $\langle n_{AZ}\rangle$ and $\langle n_{OLD}\rangle\sim\left(\frac{\sqrt{\lambda_0}}{L}\right)^{-D}$ (Fig.[\ref{fig3}]), that is the average number  of  particles composing  a DLA in a  box $L\times L$.

\noi The observed stretched exponential behavior is explained within the conformal mapping framework. The collapsing time $n_C$ can be expressed as $n_C^{-1}\propto \langle p\left(\left\{_{(1)}\Delta\vartheta_{n}^{N}\right\}\right)p\left(\left\{_{(2)}\Delta\vartheta_{n}^{N}\right\}\right)\rangle\Pi_{k=0}^{n_{AZ}}\delta\left(_{(1)}\Delta\vartheta_{N-k}^{N}-_{(2)}\Delta\vartheta_{N-k}^{N}\right)$ where the joint probability distribution  $p\left(\{_{(1)}\Delta\vartheta_{n}^{N}\}\right)$ is $ p\left(_{(1)}\Delta\vartheta_{N-n_{AZ}}^{N},\ldots,_{(1)}\Delta\vartheta_{N}^{N}\right)$, and $\delta(x)$ is the Dirac delta function. Assuming $\Delta\vartheta_{n}^{N}$ uncorrelated, we have that $n_C^{-1}\propto \Pi_{k=0}^{n_{AZ}}\langle p\left(_{(1)}\Delta\vartheta_{N-k}^{N}\right)^2\rangle$. Now, taking $p\left(_{(1)}\Delta\vartheta_{N-k}^{N}\right)\sim e^{-\frac{_{(1)}\Delta\vartheta_{N-k}^{N}}{\langle _{(1)}\Delta\vartheta_{N-k}^{N}\rangle}}$, thanks to (\ref{measure_map}) we finally obtain $n_C\sim \Pi_{k=0}^{n_{AZ}} \langle e^{2\frac{P^{(N)}(z_{N-k})}{\mu_{N,N-k}}}\rangle\sim\Pi_{k=0}^{n_{AZ}} \langle e^{2P^{(N)}(z_{N-k})\left(\frac{\sqrt\lambda_0}{L}\right)^{-\alpha\left(\frac{k}{\langle n_P\rangle}\right)}}\rangle$

{\em Conclusions--} We have shown that the harmonic measure is  stationary, unique and invariant on the DLA interface. As a matter of fact, within this comprehensive framework, the system's stationarity entails that   multiscaling, multifractality and conformal invariance appear as a unique emergent property of the growth dynamics. Moreover the stationarity allows the precise definition  of characteristic times, whose scaling exhibit a sole critical exponent: the aggregate's fractal dimension. This is at odds with radial DLAs, for which a stationary phase and an ensuing single scaling exponent cannot be identified, casting very fundamental doubts on the possible existence and definition of a fractal dimension in this geometry.  Most important, the uniqueness and invariance  of the harmonic measure  paves the way for the notion of \emph{ergodicity} in fractal growth phenomena.

The authors thank F. Martinelli for valuable suggestions. A.T. thanks F. Stenico, E. Mastrostefano and the Director of the Institute IASI-CNR for granting access to the openMosix cluster HYDRA.

\end{document}